\documentclass[12pt]{article}
\usepackage{graphicx}
\usepackage{amsmath}
\usepackage{amssymb}
\usepackage{caption2}
\begin{document}
\bibliographystyle{prsty}
\begin{center}
{\large {\bf \sc{   Structure of the $(0^+,1^+)$   mesons $B_{s0}$
and $B_{s1}$,  and the strong coupling constants $g_{B_{s0} B K}$
and
$g_{B_{s1} B^* K}$}}} \\[2mm]
Z. G. Wang \footnote{E-mail,wangzgyiti@yahoo.com.cn.  }    \\
Department of Physics, North China Electric Power University,
Baoding 071003, P. R. China
\end{center}

\begin{abstract}
In this article, we take the point of view that the bottomed
$(0^+,1^+)$  mesons $B_{s0}$ and $B_{s1}$ are the conventional
$b\bar{s}$ meson,  and  calculate the strong coupling constants
$g_{B_{s0} B K}$ and $g_{B_{s1} B^* K}$ with the light-cone QCD sum
rules.  The numerical values of strong coupling constants $g_{B_{s1}
B^* K}$ and $g_{B_{s0} B K}$ are very large, and support the
hadronic dressing mechanism. Just like the scalar mesons $f_0(980)$,
$a_0(980)$, $D_{s0}$ and axial-vector meson $D_{s1}$, the
$(0^+,1^+)$ bottomed mesons $B_{s0}$ and $B_{s1}$ may have small
$b\bar{s}$ kernels of the typical $b\bar{s}$ meson size, the strong
couplings to the hadronic channels (or the virtual mesons loops) may
result in smaller masses than the conventional $b\bar{s}$ mesons in
the potential  quark models, and enrich the pure $b\bar{s}$ states
with other components.
\end{abstract}

PACS numbers:  12.38.Lg; 13.25.Hw; 14.40.Nd

{\bf{Key Words:}}  Bottomed mesons, light-cone QCD sum rules
\section{Introduction}

 Recently, the CDF Collaboration  reports the first observation of two narrow resonances
consistent with the orbitally excited $P$-wave $B_s$ mesons using
$1$ $\mathrm{fb^{-1}}$ of $p\overline{p}$ collisions at $\sqrt{s} =
1.96 \rm{TeV} $ collected with the CDF II detector at the Fermilab
Tevatron \cite{CDF}. The masses of the two states are $M(B_{s1})=(
5829.4 \pm 0.7) \rm{MeV}$ and $M(B_{s2}^*) = (5839.7 \pm
0.7)\rm{MeV}$, and they can be assigned as the $J^P=(1^+,2^+)$
states in the heavy quark effective theory \cite{Neubert94}. The D0
Collaboration reports the direct observation of the excited $P$-wave
state $B_{s2}^*$
 in fully reconstructed decays to $B^+K^-$, the mass of the $B_{s2}^*$ meson is measured to be
 $(5839.6 \pm 1.1  \pm 0.7) \rm{MeV}$ \cite{D0}. While the  $B_s$ states with spin-parity
 $J^P=(0^+,1^+)$ are still lack experimental evidence.

The masses of the  $B_s$ mesons with $(0^+,1^+)$ have been estimated
with the potential quark models, heavy quark effective theory and
lattice QCD
\cite{BBmeson1,BBmeson2,BBmeson3,BBmeson4,BBmeson5,BBmeson6,BBmeson7,BBmeson8,BBmeson9,BBmeson10,Simonov07,Matsuki05,Matsuki07},
the values are different from each other.
 In our previous work \cite{Wang0712}, we study the masses of the bottomed $(0^+,1^+)$
  mesons with the QCD sum rules,  and observe that the central values
are below the corresponding $BK$ and $B^*K$ thresholds respectively.
The strong decays $B_{s0}\rightarrow BK$ and $B_{s1}\rightarrow
B^*K$ are kinematically  forbidden, the $P$-wave heavy mesons
$B_{s0}$ and $B_{s1}$ can decay through the isospin violation
precesses $B_{s0}\rightarrow B_s\eta\rightarrow B_s\pi^0$ and
$B_{s1}\rightarrow B_s^*\eta\rightarrow
 B_s^*\pi^0$, respectively \cite{Wang0801}. The   $\eta$ and $\pi^0$ transition matrix is very small according to
   Dashen's
 theorem \cite{Dashen}$,
 t_{\eta\pi} = \langle \pi^0 |\mathcal {H}
 |\eta\rangle=-0.003\rm{GeV}$, they
maybe very narrow. The  bottomed  mesons $B_{s0}$ and $B_{s1}$ may
have interesting feature, just like their charmed cousins $D_{s0}$
and $D_{s1}$, have small
  $b\bar{s}$  kernels of the typical
$b\bar{s}$  mesons size, strong couplings to the virtual
intermediate hadronic states (or the virtual mesons loops) may
result in smaller masses than the conventional  $b\bar{s}$ mesons in
the potential quark models, enrich the pure   $b\bar{s}$ states with
other components
\cite{Simonov07,Swanson06R,Colangelo04R,Wang23,Wang24,Wang06}.

In previous works,  the  mesons $f_0(980)$, $a_0(980)$, $D_{s0}$ and
$D_{s1}$ are taken as the conventional $q\bar{q}$ and $c\bar{s}$
states respectively, and the values of the strong coupling constants
$g_{f_0KK}$, $g_{a_0KK}$, $g_{D_{s0} DK}$ and $g_{D_{s1} D^*K}$ are
calculated with the light-cone QCD sum rules
\cite{Wang23,Wang24,Wang06,Colangelo03,Wang04}. The large values of
the strong coupling constants support the hadronic dressing
mechanism.

In this article, we take the bottomed mesons $B_{s0}$ and $B_{s1}$
as the conventional $b\bar{s}$ states, and calculate the values of
the strong coupling constants $g_{B_{s0} BK}$ and $g_{B_{s1} B^*K}$
with the light-cone QCD sum rules, and study the possibility of the
hadronic dressing mechanism in the bottomed channels.

The light-cone QCD sum rules approach carries out the operator
product expansion near the light-cone $x^2\approx 0$ instead of the
short distance $x\approx 0$ while the non-perturbative matrix
elements are parameterized by the light-cone distribution amplitudes
(which classified according to their twists)  instead of
 the vacuum condensates \cite{LCSR1,LCSR2,LCSR3,LCSR4,LCSR5,LCSRreview}. The non-perturbative
 parameters in the light-cone distribution amplitudes are calculated by
 the conventional QCD  sum rules
 and the  values are universal \cite{SVZ791,SVZ792,Reinders85,Narison89}.

The article is arranged as: in Section 2, we derive the strong
coupling constants  $g_{B_{s0} B K}$ and  $g_{B_{s1} B^* K}$ with
the light-cone QCD sum rules; in Section 3, the numerical result and
discussion; and in Section 4, conclusion.

\section{Strong coupling constants  $g_{B_{s1} B^* K}$ and $g_{B_{s0} B K}$ with light-cone QCD sum rules}

In the following, we write down the definition  for the strong
coupling constants  $g_{B_{s0} B K}$ and  $g_{B_{s1} B^* K}$,
\begin{eqnarray}
\langle B_{s1}|B^* K\rangle&=&-ig_{B_{s1} B^* K}\eta^* \cdot
\epsilon=-iM_{A}\hat{g}_{B_{s1} B^* K}
\eta^* \cdot \epsilon \, \, , \nonumber\\
\langle B_{s0}| BK\rangle&=&g_{B_{s0} BK}=M_{S}\hat{g}_{B_{s0} B K}
\, \, ,
\end{eqnarray}
where the $\epsilon_\alpha$ and $\eta_\alpha$ are the polarization
vectors of the mesons $B^*$ and $B_{s1}$ respectively. The masses
 $M_{S}$ and $M_{A}$ can serve as an energy scale, we
factorize the masses from the corresponding strong coupling
constants $g_{B_{s0} B K}$ and  $g_{B_{s1} B^* K}$ respectively.

We study the strong coupling constants  with the
 two-point correlation functions $\Pi_{\mu\nu}(p,q)$ and $\Pi_{\mu}(p,q)$ respectively,
\begin{eqnarray}
\Pi_{\mu \nu}(p,q)&=&i \int d^4x \, e^{-i q \cdot x} \,
\langle 0 |T\left\{J^V_\mu(0) J_{\nu}^{A+}(x)\right\}|K(p)\rangle \, , \\
\Pi_{\mu }(p,q)&=&i \int d^4x \, e^{-i q \cdot x} \,
\langle 0 |T\left\{J^5_{\mu}(0) J^{S+}(x)\right\}|K(p)\rangle \, , \\
J^V_\mu(x)&=&{\bar u}(x)\gamma_\mu  b(x)\, , \nonumber \\
J^A_{\mu}(x)&=&{\bar s}(x)\gamma_\mu \gamma_5 b(x)\, , \nonumber \\
J^5_{\mu}(x)&=&{\bar u}(x)\gamma_\mu \gamma_5 b(x)\, ,\nonumber  \\
J^S(x)&=&{\bar s}(x) b(x)\, ,
\end{eqnarray}
where the currents $J^V_\mu(x)$, $J^A_\mu(x)$, $J^5_\mu(x)$ and
$J^S(x)$ interpolate the bottomed mesons $B^*$, $B_{s1}$, $B$ and
$B_{s0}$,  respectively, the external $K$ meson has four momentum
$p_\mu$ with $p^2=m_K^2$. The correlation functions
$\Pi_{\mu\nu}(p,q)$ and $\Pi_{\mu}(p,q)$ can be decomposed as
\begin{eqnarray}
\Pi_{\mu \nu}(p,q)&=&i \Pi_A(p,q) g_{\mu\nu}+\Pi_{A1}(p,q)(p_\mu
q_\nu+p_\nu q_\mu)
+\cdots  \, , \nonumber\\
\Pi_{\mu}(p,q)&=&i \Pi_S(p,q) q_{\mu}+\Pi_{S1}(p,q)p_\mu +\cdots
\end{eqnarray}
due to the Lorentz invariance.

According to the basic assumption of current-hadron duality in the
QCD sum rules approach \cite{SVZ791,SVZ792,Reinders85,Narison89}, we
can insert  a complete series of intermediate states with the same
quantum numbers as the current operators $J^V_\mu(x)$, $J^A_\mu(x)$,
$J^5_\mu(x)$ and $J^S(x)$  into the correlation functions
$\Pi_{\mu\nu}(p,q)$ and $\Pi_{\mu}(p,q)$ to obtain the hadronic
representation. After isolating the ground state contributions from
the pole terms of the mesons $B^*$, $B_{s1}$, $B$ and $B_{s0}$, we
get the following results,
\begin{eqnarray}
\Pi_{\mu\nu}&=&\frac{\langle0| J^V_{\mu}(0)\mid
B^*(q+p)\rangle\langle B^*| B_{s1}K\rangle  \langle
B_{s1}(q)|J_{\nu}^{A+}(0)| 0\rangle}
  {\left[M_{B^*}^2-(q+p)^2\right]\left[M_{A}^2-q^2\right]}  + \cdots \nonumber \\
&=&-\frac{i g_{B_{s1}B^*K} f_{B^*} f_{A}  M_{B^*}M_{A}}
  {\left[M_{B^*}^2-(q+p)^2\right]\left[M_{A}^2-q^2\right]}g_{\mu\nu} + \cdots
  , \\
  \Pi_{\mu}&=&\frac{\langle0| J^5_{\mu}(0)\mid B(q+p)\rangle\langle
B| B_{s0}K\rangle  \langle B_{s0}(q)|J^{S+}(0)| 0\rangle}
  {\left[M_{B}^2-(q+p)^2\right]\left[M_{S}^2-q^2\right]}  + \cdots \nonumber \\
&=&\frac{i g_{B_{s0}BK} f_{B} f_{S}  M_{S}}
  {\left[M_{B}^2-(q+p)^2\right]\left[M_{S}^2-q^2\right]}(p+q)_{\mu} + \cdots
  ,
\end{eqnarray}
where the following definitions for the weak decay constants have
been used,
\begin{eqnarray}
\langle0 | J^V_{\mu}(0)|B^*(p)\rangle&=&f_{B^*}M_{B^*}\epsilon_\mu\,, \nonumber\\
\langle0|J^A_{\mu}(0)|B_{s1}(p)\rangle&=&f_{A}M_{A}\eta_\mu\,,\nonumber \\
\langle0 | J^5_{\mu}(0)|B(p)\rangle&=&if_{B}p_\mu\,, \nonumber\\
\langle0 | J^S(0)|B_{s0}(p)\rangle&=&f_{S}M_{S} \, .
\end{eqnarray}

In the following, we briefly outline the  operator product expansion
for the correlation functions $\Pi_{\mu \nu}(p,q)$ and $\Pi_{\mu
}(p,q)$ in perturbative QCD theory. The calculations are performed
at the large space-like momentum regions $(q+p)^2\ll 0$ and  $q^2\ll
0$, which correspond to the small light-cone distance $x^2\approx 0$
required by the validity of the operator product expansion approach.
We write down the propagator of a massive quark in the external
gluon field in the Fock-Schwinger gauge firstly \cite{Belyaev94},
\begin{eqnarray}
\langle 0 | T \{q_i(x_1)\, \bar{q}_j(x_2)\}| 0 \rangle &=&
 i \int\frac{d^4k}{(2\pi)^4}e^{-ik(x_1-x_2)}\nonumber\\
 && \left\{
\frac{\not\!k +m}{k^2-m^2} \delta_{ij} -\int\limits_0^1 dv\, g_s \,
G^{\mu\nu}_{ij}(vx_1+(1-v)x_2) \right. \nonumber \\
&&\left. \Big[ \frac12 \frac {\not\!k
+m}{(k^2-m^2)^2}\sigma_{\mu\nu} - \frac1{k^2-m^2}v(x_1-x_2)_\mu
\gamma_\nu \Big]\right\}\, .
\end{eqnarray}
 Substituting
the above $b$ quark propagator and the corresponding $K$ meson
light-cone distribution amplitudes into the correlation functions
$\Pi_{\mu\nu}(p,q)$ and $\Pi_{\mu}(p,q)$,  and completing the
integrals over the variables $x$ and $k$, finally we obtain the
analytical results, which are given explicitly in the appendix.

In calculation, the  two-particle and three-particle $K$ meson
light-cone distribution amplitudes have been used
\cite{PSLC1,PSLC2,PSLC3,PSLC4}, the explicit expressions are given
in the appendix. The parameters in the light-cone distribution
amplitudes are scale dependent and are estimated with the QCD sum
rules \cite{PSLC1,PSLC2,PSLC3,PSLC4}. In this article, the energy
scale $\mu$ is chosen to be $\mu=1\rm{GeV}$, to be more precise, one
can choose $\mu=\sqrt{M_B^2-m_b^2}\approx 2.4\rm{GeV}$.

After straightforward calculations, we obtain the final expressions
of the double Borel transformed correlation functions $\Pi_A
(M_1^2,M_2^2)$ and $\Pi_S (M_1^2,M_2^2)$ at the level of quark-gluon
degrees of freedom. The masses of  the bottomed mesons are
$M_{A}=5.72\rm{GeV}$, $M_{S}=5.70\rm{GeV}$, $M_{B^*}=5.33\rm{GeV}$
and $M_{B}=5.28\rm{GeV}$,
\begin{eqnarray}
 \frac{M_{A}^2}{M_{A}^2+M_{B^*}^2}\approx
\frac{M_{S}^2}{M_{S}^2+M_{B}^2}\approx0.54 \, ,
\end{eqnarray}
 there exists an overlapping working window for the two Borel
parameters $M_1^2$ and $M_2^2$, it's convenient to take the value
$M_1^2=M_2^2$. We introduce the threshold parameter $s_0$ and make
the simple replacement,
\begin{eqnarray}
e^{-\frac{m_b^2+u_0(1-u_0)m_K^2}{M^2}} \rightarrow
e^{-\frac{m_b^2+u_0(1-u_0)m_K^2}{M^2} }-e^{-\frac{s_0}{M^2}}
\nonumber
\end{eqnarray}
 to subtract the contributions from the high resonances  and
  continuum states \cite{Belyaev94}, finally we obtain the sum rules  for the strong coupling
  constants $g_{B_{s0}BK}$ and
$g_{B_{s1}B^*K}$,

\begin{eqnarray}
g_{B_{s0}BK}&=& \frac{1}{f_{B}f_{S}M_{S}}\exp\left(
\frac{M^2_{S}}{M_1^2} +\frac{M^2_{B}}{M_2^2}
\right)\left\{\left[\exp\left(- \frac{\Xi}{M^2}\right)-\exp\left(-
\frac{s_S^0}{M^2}\right)\right]  \right.\nonumber\\
&& \frac{f_K
m_K^2M^2}{m_u+m_s}\left[\varphi_p(u_0)-\frac{d\varphi_\sigma(u_0)}{6du_0}\right]
+\exp\left(-\frac{\Xi}{M^2}\right)\left[ -m_bf_K m_K^2  \int_0^{u_0}
dt B(t)  \right.\nonumber\\
  &&+f_{3K}m_K^2 \int_0^{u_0}d\alpha_s
  \int_{u_0-\alpha_s}^{1-\alpha_s}d\alpha_g
\varphi_{3K}(1-\alpha_s-\alpha_g,\alpha_g,\alpha_s)\frac{2(\alpha_s+\alpha_g-u_0)-3\alpha_g
}{\alpha_g^2}
\nonumber\\
&&-\frac{2m_bf_K m_K^4}{M^2}  \int_{1-u_0}^1 d\alpha_g
\frac{1-u_0}{\alpha_g^2}\int_0^{\alpha_g}
d\beta\int_0^{1-\beta}d\alpha \Phi(1-\alpha-\beta,\beta,\alpha)
\nonumber \\
&& +\frac{2m_bf_K m_K^4}{M^2}\left(\int_0^{1-u_0} d\alpha_g
\int^{u_0}_{u_0-\alpha_g} d\alpha_s \int_0^{\alpha_s} d\alpha
+\int^1_{1-u_0} d\alpha_g \int^{1-\alpha_g}_{u_0-\alpha_g} d\alpha_s
\int_0^{\alpha_s} d\alpha\right) \nonumber\\
&&\left.\left.\frac{\Phi(1-\alpha-\alpha_g,\alpha_g,\alpha)}{\alpha_g}
\right]\right\} \, ,
\end{eqnarray}

\begin{eqnarray}
 g_{B_{s1}B^*K}&=&
\frac{1}{f_{B^*}f_{A}M_{B^*}M_{A}} \exp\left( \frac{M^2_{A}}{M_1^2}
+\frac{M^2_{B^*}}{M_2^2} \right) \left\{\left[\exp\left(-
\frac{\Xi}{M^2}\right)-\exp\left(-
\frac{s^0_A}{M^2}\right)\right] \right.\nonumber\\
 &&f_K\left[\frac{ m_b m_K^2M^2}{m_u+m_s}
\varphi_p(u_0)
+\frac{m_K^2(M^2+m_b^2)}{8}  \frac{d}{du_0}A(u_0) -\frac{M^4}{2} \frac{d}{du_0}\phi_K(u_0)\right] \nonumber\\
&&-\exp\left(-\frac{\Xi}{M^2}\right) \left[f_K m_b^2 m_K^2
\int_0^{u_0} dt B(t) \right.\nonumber\\
&&+ m_K^2 \int_0^{u_0} d\alpha_s
\int_{u_0-\alpha_s}^{1-\alpha_s} d\alpha_g \frac{(u_0f_Km_K^2\Phi+f_{3K}m_b \varphi_{3K})(1-\alpha_s-\alpha_g,\alpha_s,\alpha_g)}{\alpha_g} \nonumber \\
&&+f_Km_K^2M^2\frac{d}{du_0}\int_0^{u_0} d\alpha_s
\int_{u_0-\alpha_s}^{1-\alpha_s} d\alpha_g
\frac{(A_\parallel-V_\parallel)(1-\alpha_s-\alpha_g,\alpha_s,\alpha_g)}{2\alpha_g}
 \nonumber \\
&&-f_Km_K^2M^2\frac{d}{du_0}\int_0^{u_0} d\alpha_s
\int_{u_0-\alpha_s}^{1-\alpha_s} d\alpha_g
A_\parallel(1-\alpha_s-\alpha_g,\alpha_s,\alpha_g)\frac{\alpha_s+\alpha_g-u_0}{\alpha_g^2}
 \nonumber \\
 &&+f_Km_K^4  \left(\int_0^{1-u_0} d\alpha_g \int^{u_0}_{u_0-\alpha_g}
d\alpha_s \int_0^{\alpha_s} d\alpha +\int^1_{1-u_0} d\alpha_g
\int^{1-\alpha_g}_{u_0-\alpha_g}
d\alpha_s \int_0^{\alpha_s} d\alpha\right)  \nonumber \\
&& \left[\frac{1}{\alpha_g}\left(3-\frac{2m_b^2}{M^2}\right)\Phi+\frac{4m_b^2}{M^2}\frac{\alpha_s+\alpha_g-u_0}{\alpha_g^2}(A_\perp+A_\parallel)\right](1-\alpha-\alpha_g,\alpha,\alpha_g)\nonumber \\
 &&-f_Km_K^4  u_0\frac{d}{du_0}\left(\int_0^{1-u_0} d\alpha_g \int^{u_0}_{u_0-\alpha_g}
d\alpha_s \int_0^{\alpha_s} d\alpha +\int^1_{1-u_0} d\alpha_g
\int^{1-\alpha_g}_{u_0-\alpha_g}
d\alpha_s \int_0^{\alpha_s} d\alpha\right)  \nonumber \\
&& \frac{\Phi(1-\alpha-\alpha_g,\alpha,\alpha_g)}{\alpha_g} \nonumber \\
&&-f_Km_K^4   \int_{1-u_0}^1 d\alpha_g \int_0^{\alpha_g} d\beta
\int_0^{1-\beta} d\alpha \left[\Phi(1-\alpha-\beta,\alpha,\beta)
\frac{1-u_0}{\alpha_g^2}
 \left(4-\frac{2m_b^2}{M^2}\right) \right.\nonumber \\
&&\left. +\frac{4m_b^2}{M^2}\frac{(1-u_0)^2}{\alpha_g^3}(A_\parallel+A_\perp)(1-\alpha-\beta,\alpha,\beta)\right]\nonumber \\
&&\left.\left.+f_Km_K^4  \frac{d}{du_0} \int_{1-u_0}^1 d\alpha_g
\int_0^{\alpha_g} d\beta \int_0^{1-\beta} d\alpha
\Phi(1-\alpha-\beta,\alpha,\beta) \frac{u_0(1-u_0)}{\alpha_g^2}
\right]\right\} ,
\end{eqnarray}
where
\begin{eqnarray}
\Xi&=&m_b^2+u_0(1-u_0)m_K^2 \, ,\nonumber \\
u_0&=&\frac{M_1^2}{M_1^2+M_2^2}\, , \nonumber \\
M^2&=&\frac{M_1^2M_2^2}{M_1^2+M_2^2} \, .
\end{eqnarray}
 The term proportional to the $M^4\frac{d}{du_0}\phi_K(u_0)$ in Eq.(12)
 depends heavily on the asymmetric
  coefficient  $a_1(\mu)$ of the twist-2 light-cone distribution amplitude
  $\phi_K(u)$ in the  limit  $u_0=\frac{1}{2}$(see also the sum rules for the strong coupling
  constant $g_{D_{s1}D^*K}$ in Ref. \cite{Wang24}), if we take the value $a_1(\mu)=0.06\pm 0.03$ \cite{PSLC1,PSLC2,PSLC3,PSLC4},
  no stable sum rules can be obtained, the value of the $g_{B_{s1}B^*K}$ changes
  significantly with the variation of the Borel
  parameter $M^2$.  In this article, we take the assumption that the $u$ and $s$ quarks
  have  symmetric momentum distributions and  neglect the coefficient $a_1(\mu)$.

In the heavy quark limit  $m_b\rightarrow \infty$,
\begin{eqnarray}
s_S^0&\rightarrow&m_b^2+2m_b\omega^0_S \, ,\nonumber\\
s_A^0&\rightarrow&m_b^2+2m_b\omega^0_A \, ,\nonumber\\
M^2_1&\rightarrow&2m_bT_1\, ,\nonumber\\
M^2_2&\rightarrow&2m_bT_2\, ,\nonumber\\
M^2&\rightarrow&2m_bT \, ,\nonumber\\
M_S&\rightarrow&m_b+\Lambda_1 \, ,\nonumber\\
M_A&\rightarrow&m_b+\Lambda_1\, ,\nonumber\\
M_B&\rightarrow&m_b+\Lambda_0\, ,\nonumber\\
M_{B^*}&\rightarrow&m_b+\Lambda_0 \, ,
\end{eqnarray}
the two sum rules in Eqs.(11-12) are reduced to the following form,
\begin{eqnarray}
g_{B_{s0}BK}&=& \frac{1}{f_{B}f_{S}}\exp\left(
\frac{\Lambda_{1}}{T_1} +\frac{\Lambda_{0}}{T_2}
\right)\left\{\left[1-\exp\left(-
\frac{\omega_S^0}{T}\right)\right]  \right.\nonumber\\
&& \left.\frac{2f_K m_K^2
T}{m_u+m_s}\left[\varphi_p(u_0)-\frac{d\varphi_\sigma(u_0)}{6du_0}\right]
 -f_K m_K^2  \int_0^{u_0}
dt B(t)  \right\} \, , \\
 g_{B_{s1}B^*K}&=&
\frac{1}{f_{B^*}f_{A}} \exp\left( \frac{\Lambda_{1}}{T_1}
+\frac{\Lambda_{0}}{T_2} \right) \left\{\left[1-\exp\left(-
\frac{\omega^0_A}{T}\right)\right] \right.\nonumber\\
 &&f_K\left[\frac{2  m_K^2 T}{m_u+m_s}
\varphi_p(u_0)
+\frac{m_K^2(2T+m_b)}{8m_b}  \frac{d}{du_0}A(u_0) -2T^2 \frac{d}{du_0}\phi_K(u_0)\right] \nonumber\\
&&- \left.f_K  m_K^2 \int_0^{u_0} dt B(t) \right\} ,
\end{eqnarray}
where the decay constants take the behavior
$f_A=\frac{C_1}{\sqrt{M_A}}$, $f_S=\frac{C_1}{\sqrt{M_S}}$,
$f_B=\frac{C_2}{\sqrt{M_B}}$ and
$f_{B^*}=\frac{C_2}{\sqrt{M_{B^*}}}$ (according to the definition in
Eq.(8)), the $C_i$ are some constants.

\section{Numerical result and discussion}
The input parameters are taken as $m_s=(140\pm 10 )\rm{MeV}$,
$m_u=(5.6\pm1.6)\rm{MeV}$, $m_b=(4.7\pm 0.1)\rm{GeV}$,
$\lambda_3=1.6\pm0.4$, $f_{3K}=(0.45\pm0.15)\times
10^{-2}\rm{GeV}^2$, $\omega_3=-1.2\pm0.7$, $\eta_4=0.6 \pm0.2 $,
$\omega_4=0.2\pm0.1$, $a_2=0.25\pm 0.15$
\cite{LCSRreview,PSLC1,PSLC2,PSLC3,PSLC4}, $f_K=0.160\rm{GeV}$,
$m_K=0.498\rm{GeV}$, $M_B=5.279\rm{GeV}$, $M_{B^*}=5.325\rm{GeV}$
\cite{PDG}, $M_{S}=(5.70\pm0.11)\rm{GeV} $,
$M_{A}=(5.72\pm0.09)\rm{GeV}$, $f_{S}= f_{A}=(0.24\pm 0.02)\rm{GeV}$
\cite{Wang0712}, $f_{B^*}=f_B=(0.17\pm0.02)\rm{GeV}$
\cite{LCSRreview,Wang04BS,Velasco07},
  $s^0_S=(37 \pm 1)\rm{GeV}^2$, $s^0_A=(38 \pm 1)\rm{GeV}^2$
\cite{Wang0712},
$\Lambda_0=\frac{M_B+3M_{B^*}}{4}-m_b=(0.6\pm0.1)\rm{GeV}$,
$\Lambda_1=\frac{M_S+3M_{A}}{4}-m_b=(1.0\pm0.1)\rm{GeV}$,
$\omega_S^0=(1.6\pm0.1)\rm{GeV}$ and
$\omega_A^0=(1.6\pm0.1)\rm{GeV}$.
  The Borel parameters are chosen as $M^2=(5-7)\rm{GeV}^2$, in this region, the
values of the strong coupling constants  $g_{B_{s1} B^*K}$ and
$g_{B_{s0} BK}$ are rather stable, which are shown in Fig.1.
\begin{figure} \centering
  \includegraphics[totalheight=6cm,width=7cm]{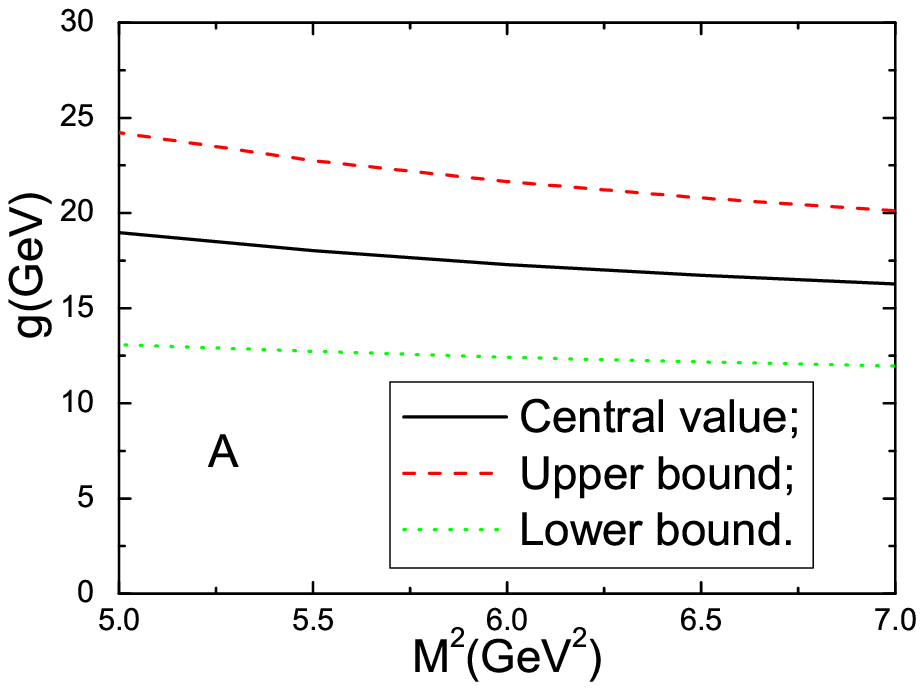}
 \includegraphics[totalheight=6cm,width=7cm]{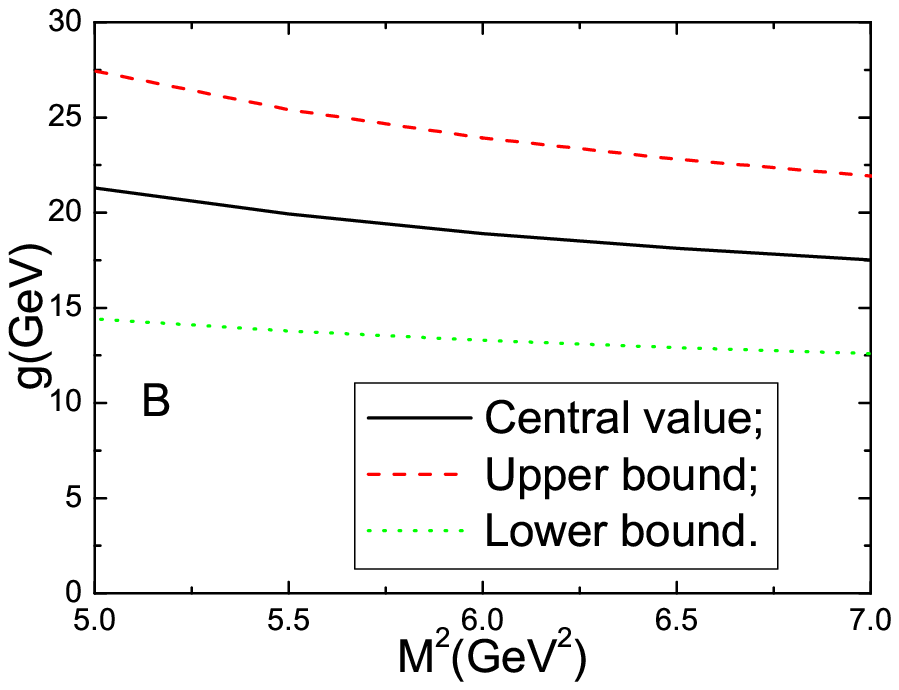}
     \caption{The strong coupling constants  $g_{B_{s1}B^*K}$(A) and  $g_{B_{s0}BK}$(B) with the Borel parameter $M^2$. }
\end{figure}
In the heavy quark limit, the Borel parameters are chosen as
$T=(0.7-1.5)\rm{GeV}$, in this region, the values of the strong
coupling constants  $g_{B_{s1} B^*K}$ and $g_{B_{s0} BK}$ are rather
stable, which are shown in Fig.2.

 In the limit of large Borel parameter $M^2$, the strong coupling
constants $g_{B_{s1} B^*K}$ and $g_{B_{s0} BK}$ take up the
following behaviors,
\begin{eqnarray}
g_{B_{s0}BK}&\propto& \frac{ M^2\varphi_p(u_0)}{f_{B}f_{S}}
\, , \nonumber \\
g_{B_{s1}B^*K}&\propto& \frac{m_b M^2\varphi_p(u_0)}{f_{B^*}f_{A}}
\, .
\end{eqnarray}
It is not unexpected, the contributions  from the two-particle
twist-3 light-cone distribution amplitude $\varphi_p(u)$ are greatly
enhanced by the large Borel parameter $M^2$,  (large) uncertainties
of the relevant parameters presented in above equations have
significant impact on the numerical results. The contributions from
the two-particle twist-2, twist-3 and twist-4 light-cone
distribution amplitudes $\phi_K(u_0)$, $\varphi_\sigma(u_0)$ and
$A(u_0)$ are zero due to symmetry property.

Taking into account all the uncertainties of the input parameters,
finally we obtain the numerical values of the strong coupling
constants
\begin{eqnarray}
  g_{B_{s1}B^*K} &=&(18.1\pm6.1) \rm{GeV} \, , \nonumber\\
  g_{B_{s0}BK} &=&(20.0\pm7.4) \rm{GeV} \, ,\nonumber \\
 \hat{g}_{B_{s1}B^*K} &=&3.2 \pm 1.1  \, ,\nonumber \\
 \hat{g}_{B_{s0}BK} &=&3.5 \pm 1.3
\end{eqnarray}
from Eqs.(11-12) and
\begin{eqnarray}
  g_{B_{s1}B^*K}=g_{B_{s0}BK}  &=&(19.6\pm5.7) \rm{GeV}
\end{eqnarray}
from Eqs.(15-16).  The uncertainties are large, about $30\%$. The
contributions from three-particle light-cone distribution amplitudes
vanish in the heavy quark limit, the uncertainties are reduced
slightly, as the dominating contributions come from the two-particle
twist-3 light-cone distribution amplitude $\varphi_p(u)$.

The large values of the strong coupling constants $g_{B_{s1}B^*K}$
and $g_{B_{s0}BK}$ obviously support  the hadronic dressing
mechanism \cite{HDress1,HDress2,HDress3},   the scalar meson
$B_{s0}$($D_{s0}$) and axial-vector meson $B_{s1}$($D_{s1}$) can be
taken as having small scalar and axial-vector  $b\bar{s}$
($c\bar{s}$) kernels of typical meson size with large virtual
$S$-wave $BK$($DK$) and $B^*K$($D^*K$) cloud respectively.

\begin{figure} \centering
  \includegraphics[totalheight=8cm,width=10cm]{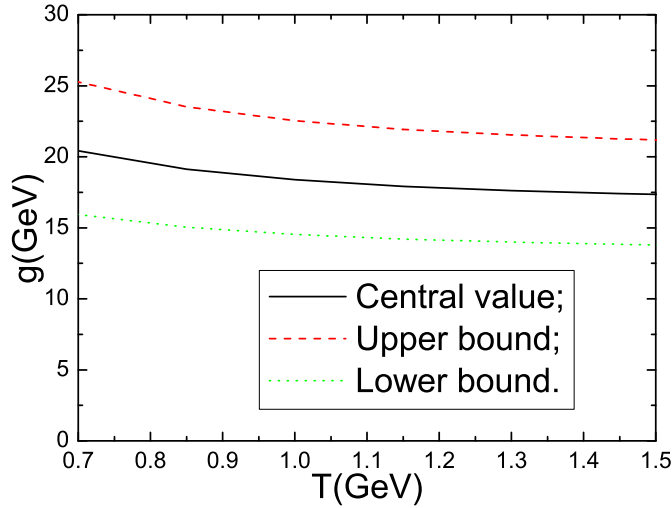}
      \caption{The strong coupling constants  $g_{B_{s1}B^*K}$ and $g_{B_{s0}BK}$ with the Borel parameter $T$ in the heavy quark limit. }
\end{figure}

\begin{table}
\begin{center}
\begin{tabular}{|c|c|c|c|c|}
\hline\hline & $g_{B_{s1}B^*K}(\rm{GeV})$& $g_{B_{s0}BK}(\rm{GeV})$ &$g_{D_{s1}D^*K}(\rm{GeV})$&$g_{D_{s0}DK}(\rm{GeV})$\\
\hline
 \cite{Wang23,Wang24} &   & & $10.5\pm3.5$&$9.3^{+2.7}_{-2.1}$\\ \hline
            \cite{Guo23,Guo24} &  23.572 &23.442&10.762&10.203\\ \hline
               This work & $18.1\pm6.1$ &$20.0\pm7.4$&&\\ \hline
This work$^*$ & $19.6\pm5.7$ &$19.6\pm5.7$&&\\ \hline
    \hline
\end{tabular}
\end{center}
\caption{ Theoretical estimations of the strong coupling constants
from different models, where $^*$ stands for the strong coupling
constants in the heavy quark limit. }
\end{table}

In Refs.\cite{Guo23,Guo24}, the authors analyze the unitarized
two-meson scattering amplitudes from the heavy-light chiral
lagrangian,  and observe that the scalar mesons $D_{s0}$ and
$B_{s0}$, and axial-vector mesons $D_{s1}$ and $B_{s1}$ appear as
the bound state poles with the strong coupling constants
$g_{D_{s0}DK}=10.203\rm{GeV}$, $g_{D_{s1}D^*K}=10.762\rm{GeV}$,
$g_{B_{s1}B^*K}=23.572\rm{GeV}$ and $g_{B_{s0}BK}=23.442\rm{GeV}$.
Our numerical results for the strong coupling constants are
certainly reasonable and can make robust predictions. However, we
take the point of view that the mesons $D_{s0}$, $B_{s0}$, $D_{s1}$
and $B_{s1}$ be bound states in the sense that they appear below the
corresponding $DK$, $BK$, $D^*K$ and $B^*K$ thresholds respectively,
their constituents may be the bare $c\bar{s}$ and $b\bar{s}$ states,
the virtual  $DK$, $BK$, $D^*K$ and $B^*K$ pairs and their mixing,
rather than the $DK$, $BK$, $D^*K$ and $B^*K$ bound states.

In Ref.\cite{Zhu06},  the authors calculate the strong coupling
constants $g_{D_{s0}D_s\eta} $ and $g_{D_{s1}D^*_s\eta} $ with the
light-cone QCD sum rules, then take into account $\eta-\pi^0$ mixing
and calculate their pionic decay widths. The bottomed mesons
 $B_{s0}$ and $B_{s1}$ can decay through the same isospin violation
mechanism, $B_{s0}\rightarrow B_s\eta\rightarrow B_s\pi^0$ and
$B_{s1}\rightarrow B_s^*\eta\rightarrow
 B_s^*\pi^0$. We study the strong coupling constants
$g_{B_{s0}B_s\eta} $ and $g_{B_{s1}B^*_s\eta}$ with the light-cone
QCD sum rules and make predictions for the corresponding small decay
widths \cite{Wang0801}.

\section{Conclusion}

In this article, we take the point of view that the  bottomed mesons
$B_{s0} $ and $B_{s1}$ are the conventional $b\bar{s}$ mesons and
calculate the strong coupling constants $g_{B_{s0} B K}$ and
$g_{B_{s1} B^* K}$ with the light-cone QCD sum rules. The numerical
results  are compatible with the existing estimations, the large
values support the hadronic dressing mechanism.    Just like the
scalar mesons $f_0(980)$, $a_0(980)$, $D_{s0}$ and axial-vector
meson $D_{s1}$, the bottomed mesons $B_{s0} $ and $B_{s1}$  may have
small $b\bar{s}$ kernels  of typical $b\bar{s}$ meson size. The
strong couplings  to virtual intermediate hadronic states (or the
virtual mesons loops) can result in  smaller masses than the
conventional $0^+$ and $1^+$ mesons in the potential quark models,
enrich the pure $b\bar{s}$ states with other components.

 \section*{Appendix}
The analytical expressions of the $\Pi_S(p,q)$ and $\Pi_A(p,q)$ at
the level of the  quark-gluon degrees of freedom,

 \begin{eqnarray}
\Pi_S&=&  \frac{f_K m_K^2}{m_u+m_s}\int_0^1du
\frac{\varphi_p(u)}{\Delta}-m_bf_K m_K^2\int_0^1du \int_0^u
dt\frac{B(t)}{\Delta^2} \nonumber\\
&&+\frac{1}{6}\frac{f_K m_K^2}{m_u+m_s}\int_0^1du
\varphi_\sigma(u)\frac{d}{du} \frac{1}{\Delta}
  \nonumber\\
  &&+f_{3K}m_K^2\int_0^1dv \int_0^1d\alpha_g
  \int_0^{1-\alpha_g}d\alpha_s
\varphi_{3K}(\alpha_u,\alpha_g,\alpha_s)\frac{2v-3
}{\Delta^2}\mid_{u=(1-v)\alpha_g+\alpha_s}
\nonumber\\
&&-4m_bf_K m_K^4\int_0^1dv v \int_0^1 d\alpha_g\int_0^{\alpha_g}
d\beta\int_0^{1-\beta}d\alpha
\frac{\Phi(1-\alpha-\beta,\beta,\alpha)}{\Delta^3}\mid_{u=1-v\alpha_g}
\nonumber \\
&& +4m_bf_K m_K^4\int_0^1 dv\int_0^1 d\alpha_g\int_0^{1-\alpha_g}
d\alpha_s
 \int_0^{\alpha_s}d\alpha
\frac{\Phi(1-\alpha-\alpha_g,\alpha_g,\alpha)}{\Delta^3}\mid_{u=(1-v)\alpha_g+\alpha_s}
\, , \nonumber \\
\Pi_A &=&-\frac{f_K m_b m_K^2}{m_u+m_s} \int_0^1 du  \frac{\varphi_p
(u)}{\Delta}+f_K m_b^2 m_K^2 \int_0^1 du \int_0^u dt \frac{B(t)}{\Delta^2} \nonumber\\
&&+\frac{f_K}{2} \int_0^1 du \left\{ \phi_K(u) \frac{d}{du}\log
\Delta+ \frac{A(u)m_K^2}{4} \frac{d}{du} \left[
\frac{1}{\Delta} +\frac{m_b^2}{\Delta^2}\right]\right\} \nonumber \\
&&+m_K^2 \int_0^1 dv \int_0^1 d\alpha_g \int_0^{1-\alpha_g}
d\alpha_s  \nonumber\\
&&\frac{\left[f_Km_K^2u\Phi+f_{3K}m_b\varphi_{3K}\right](1-\alpha_s-\alpha_g,\alpha_s,\alpha_g)}{\Delta^2} \mid_{u=\alpha_s+(1-v)\alpha_g} \nonumber \\
&&-\frac{f_Km_K^2}{2} \int_0^1 dv \int_0^1 d\alpha_g
\int_0^{1-\alpha_g}
d\alpha_s  \left[(1-2v)A_\parallel -V_\parallel\right](1-\alpha_s-\alpha_g,\alpha_s,\alpha_g) \nonumber\\
&& \frac{d}{du} \frac{1}{\Delta}\mid_{u=\alpha_s+(1-v)\alpha_g} \nonumber \\
&&+f_Km_K^4 \int_0^1 dv \int_0^1 d\alpha_g \int_0^{1-\alpha_g}
d\alpha_s \int_0^{\alpha_s} d\alpha \Phi(1-\alpha-\alpha_g,\alpha,\alpha_g)  \nonumber \\
&&\left\{ \frac{4}{\Delta^2}-\frac{4m_b^2}{\Delta^3}+u\frac{d}{du}
\frac{1}{\Delta^2} \right\}\mid_{u=\alpha_s+(1-v)\alpha_g} \nonumber \\
&&-f_Km_K^4 \int_0^1 dvv \int_0^1 d\alpha_g \int_0^{\alpha_g}
d\beta \int_0^{1-\beta} d\alpha \Phi(1-\alpha-\beta,\alpha,\beta)  \nonumber \\
&&\left\{ \frac{4}{\Delta^2}-\frac{4m_b^2}{\Delta^3}+u\frac{d}{du}
\frac{1}{\Delta^2} \right\}\mid_{u=1-v\alpha_g} \nonumber \\
&&+8f_Km_b^2m_K^4 \int_0^1 dv v\int_0^1 d\alpha_g
\int_0^{1-\alpha_g}
d\alpha_s \int_0^{\alpha_s} d\alpha(A_\parallel+A_\perp)(1-\alpha-\alpha_g,\alpha,\alpha_g)  \nonumber \\
&& \frac{1}{\Delta^3}\mid_{u=\alpha_s+(1-v)\alpha_g} \nonumber \\
&&-8f_K m_b^2m_K^4 \int_0^1 dvv^2 \int_0^1 d\alpha_g
\int_0^{\alpha_g}
d\beta \int_0^{1-\beta} d\alpha (A_\parallel+A_\perp)(1-\alpha-\beta,\alpha,\beta)  \nonumber \\
&& \frac{1}{\Delta^3} \mid_{u=1-v\alpha_g}  \,\, ,
\end{eqnarray}
where
\begin{eqnarray}
\Delta&=&m_b^2-(q+up)^2 \, ,\nonumber \\
\Phi&=&A_\parallel+A_\perp-V_\parallel-V_\perp\, .
\end{eqnarray}
 The light-cone distribution amplitudes of the $K$ meson are defined
 by
\begin{eqnarray}
\langle0| {\bar u} (0) \gamma_\mu \gamma_5 s(x) |K(p)\rangle& =& i
f_K p_\mu \int_0^1 du  e^{-i u p\cdot x}
\left\{\phi_K(u)+\frac{m_K^2x^2}{16}
A(u)\right\}\nonumber\\
&&+f_K m_K^2\frac{ix_\mu}{2p\cdot x}
\int_0^1 du  e^{-i u p \cdot x} B(u) \, , \nonumber\\
\langle0| {\bar u} (0) i \gamma_5 s(x) |K(p)\rangle &=& \frac{f_K
m_K^2}{
m_s+m_u}\int_0^1 du  e^{-i u p \cdot x} \varphi_p(u)  \, ,  \nonumber\\
\langle0| {\bar u} (0) \sigma_{\mu \nu} \gamma_5 s(x) |K(p)\rangle
&=&i(p_\mu x_\nu-p_\nu x_\mu)  \frac{f_K m_K^2}{6 (m_s+m_u)}
\int_0^1 du
e^{-i u p \cdot x} \varphi_\sigma(u) \, ,  \nonumber\\
\langle0| {\bar u} (0) \sigma_{\alpha \beta} \gamma_5 g_s G_{\mu
\nu}(v x)s(x) |K(p)\rangle&=& f_{3 K}\left\{(p_\mu p_\alpha
g^\bot_{\nu
\beta}-p_\nu p_\alpha g^\bot_{\mu \beta}) -(p_\mu p_\beta g^\bot_{\nu \alpha}\right.\nonumber\\
&&\left.-p_\nu p_\beta g^\bot_{\mu \alpha})\right\} \int {\cal
D}\alpha_i \varphi_{3 K} (\alpha_i)
e^{-ip \cdot x(\alpha_s+v \alpha_g)} \, ,\nonumber\\
\langle0| {\bar u} (0) \gamma_{\mu} \gamma_5 g_s G_{\alpha
\beta}(vx)s(x) |K(p)\rangle&=&  f_Km_K^2p_\mu  \frac{p_\alpha
x_\beta-p_\beta x_\alpha}{p
\cdot x}\nonumber\\
&&\int{\cal D}\alpha_i A_{\parallel}(\alpha_i) e^{-ip\cdot
x(\alpha_s +v \alpha_g)}\nonumber \\
&&+ f_Km_K^2 (p_\beta g_{\alpha\mu}-p_\alpha
g_{\beta\mu})\nonumber\\
&&\int{\cal D}\alpha_i A_{\perp}(\alpha_i)
e^{-ip\cdot x(\alpha_s +v \alpha_g)} \, ,  \nonumber\\
\langle0| {\bar u} (0) \gamma_{\mu}  g_s \tilde G_{\alpha
\beta}(vx)s(x) |K(p)\rangle&=&f_Km_K^2 p_\mu  \frac{p_\alpha
x_\beta-p_\beta x_\alpha}{p \cdot
x}\nonumber\\
&&\int{\cal D}\alpha_i V_{\parallel}(\alpha_i) e^{-ip\cdot
x(\alpha_s +v \alpha_g)}\nonumber \\
&&+ f_Km_K^2 (p_\beta g_{\alpha\mu}-p_\alpha
g_{\beta\mu})\nonumber\\
&&\int{\cal D}\alpha_i V_{\perp}(\alpha_i) e^{-ip\cdot x(\alpha_s +v
\alpha_g)} \, ,
\end{eqnarray}
where the operator $\tilde G_{\alpha \beta}$  is the dual of the
$G_{\alpha \beta}$, $\tilde G_{\alpha \beta}= {1\over 2}
\epsilon_{\alpha \beta  \mu\nu} G^{\mu\nu} $ and ${\cal{D}}\alpha_i$
is defined as ${\cal{D}} \alpha_i =d \alpha_1 d \alpha_2 d \alpha_3
\delta(1-\alpha_1 -\alpha_2 -\alpha_3)$. The  light-cone
distribution amplitudes are parameterized as
\begin{eqnarray}
\phi_K(u,\mu)&=&6u(1-u)
\left\{1+a_1C^{\frac{3}{2}}_1(2u-1)+a_2C^{\frac{3}{2}}_2(2u-1)
\right\}\, , \nonumber\\
\varphi_p(u,\mu)&=&1+\left\{30\eta_3-\frac{5}{2}\rho^2\right\}C_2^{\frac{1}{2}}(2u-1)\nonumber \\
&&+\left\{-3\eta_3\omega_3-\frac{27}{20}\rho^2-\frac{81}{10}\rho^2 a_2\right\}C_4^{\frac{1}{2}}(2u-1)\, ,  \nonumber \\
\varphi_\sigma(u,\mu)&=&6u(1-u)\left\{1
+\left[5\eta_3-\frac{1}{2}\eta_3\omega_3-\frac{7}{20}\rho^2-\frac{3}{5}\rho^2 a_2\right]C_2^{\frac{3}{2}}(2u-1)\right\}\, , \nonumber \\
\varphi_{3K}(\alpha_i,\mu) &=& 360 \alpha_u \alpha_s \alpha_g^2
\left \{1 +\lambda_3(\alpha_u-\alpha_s)+ \omega_3 \frac{1}{2} ( 7
\alpha_g
- 3) \right\} \, , \nonumber\\
V_{\parallel}(\alpha_i,\mu) &=& 120\alpha_u \alpha_s \alpha_g \left(
v_{00}+v_{10}(3\alpha_g-1)\right)\, ,
\nonumber \\
A_{\parallel}(\alpha_i,\mu) &=& 120 \alpha_u \alpha_s \alpha_g
a_{10} (\alpha_s-\alpha_u)\, ,
\nonumber\\
V_{\perp}(\alpha_i,\mu) &=& -30\alpha_g^2
\left\{h_{00}(1-\alpha_g)+h_{01}\left[\alpha_g(1-\alpha_g)-6\alpha_u
\alpha_s\right] \right.  \nonumber\\
&&\left. +h_{10}\left[
\alpha_g(1-\alpha_g)-\frac{3}{2}\left(\alpha_u^2+\alpha_s^2\right)\right]\right\}\,
, \nonumber\\
A_{\perp}(\alpha_i,\mu) &=&  30 \alpha_g^2 (\alpha_u-\alpha_s) \left\{h_{00}+h_{01}\alpha_g+\frac{1}{2}h_{10}(5\alpha_g-3)  \right\}, \nonumber\\
A(u,\mu)&=&6u(1-u)\left\{
\frac{16}{15}+\frac{24}{35}a_2+20\eta_3+\frac{20}{9}\eta_4 \right.
\nonumber \\
&&+\left[
-\frac{1}{15}+\frac{1}{16}-\frac{7}{27}\eta_3\omega_3-\frac{10}{27}\eta_4\right]C^{\frac{3}{2}}_2(2u-1)
\nonumber\\
&&\left.+\left[
-\frac{11}{210}a_2-\frac{4}{135}\eta_3\omega_3\right]C^{\frac{3}{2}}_4(2u-1)\right\}+\left\{
 -\frac{18}{5}a_2+21\eta_4\omega_4\right\} \nonumber\\
 && \left\{2u^3(10-15u+6u^2) \log u+2\bar{u}^3(10-15\bar{u}+6\bar{u}^2) \log \bar{u}
 \right. \nonumber\\
 &&\left. +u\bar{u}(2+13u\bar{u})\right\} \, ,\nonumber\\
 g_K(u,\mu)&=&1+g_2C^{\frac{1}{2}}_2(2u-1)+g_4C^{\frac{1}{2}}_4(2u-1)\, ,\nonumber\\
 B(u,\mu)&=&g_K(u,\mu)-\phi_K(u,\mu)\, ,
\end{eqnarray}
where
\begin{eqnarray}
h_{00}&=&v_{00}=-\frac{\eta_4}{3} \, ,\nonumber\\
a_{10}&=&\frac{21}{8}\eta_4 \omega_4-\frac{9}{20}a_2 \, ,\nonumber\\
v_{10}&=&\frac{21}{8}\eta_4 \omega_4 \, ,\nonumber\\
h_{01}&=&\frac{7}{4}\eta_4\omega_4-\frac{3}{20}a_2 \, ,\nonumber\\
h_{10}&=&\frac{7}{2}\eta_4\omega_4+\frac{3}{20}a_2 \, ,\nonumber\\
g_2&=&1+\frac{18}{7}a_2+60\eta_3+\frac{20}{3}\eta_4 \, ,\nonumber\\
g_4&=&-\frac{9}{28}a_2-6\eta_3\omega_3 \, ,
\end{eqnarray}
 here  $ C_2^{\frac{1}{2}}$, $ C_4^{\frac{1}{2}}$
 and $ C_2^{\frac{3}{2}}$ are Gegenbauer polynomials,
  $\eta_3=\frac{f_{3K}}{f_K}\frac{m_q+m_s}{m_K^2}$ and  $\rho^2={(m_u+m_s)^2\over m_K^2}$
 \cite{LCSR1,LCSR2,LCSR3,LCSR4,PSLC1,PSLC2,PSLC3,PSLC4}.

\section*{Acknowledgments}
This  work is supported by National Natural Science Foundation,
Grant Number 10405009, 10775051, and Program for New Century
Excellent Talents in University, Grant Number NCET-07-0282, and Key
Program Foundation of NCEPU.


\begin{thebibliography}{99}



\bibitem{CDF}  T. Aaltonen, et al, arXiv:0710.4199.
\bibitem{Neubert94} M. Neubert, Phys. Rept. {\bf 245} (1994) 259.
\bibitem{D0} V. Abazov, et al, arXiv:0711.0319.

\bibitem{BBmeson1} D. Ebert, V. O. Galkin and R. N. Faustov, Phys. Rev. {\bf D57} (1998)
5663.

\bibitem{BBmeson2} S. Godfrey and R. Kokoski, Phys. Rev. {\bf D43} (1991)
1679.

\bibitem{BBmeson3} W. A. Bardeen, E. J. Eichten and C. T. Hill, Phys. Rev. {\bf D68}
(2003) 054024.

\bibitem{BBmeson4} P. Colangelo, F. De Fazio and R. Ferrandes, Nucl. Phys. Proc. Suppl.
{\bf 163} (2007) 177.

\bibitem{BBmeson5} A. M. Green, et al, Phys. Rev. {\bf D69}
(2004) 094505.

\bibitem{BBmeson6} M. Di Pierro and E. Eichten, Phys. Rev. {\bf D64} (2001)
114004.

\bibitem{BBmeson7} J. Vijande, A. Valcarce and F. Fernandez,
arXiv:0711.2359.

\bibitem{BBmeson8} M. A. Nowak, M. Rho and I. Zahed, Acta. Phys. Polon. {\bf B35}
(2004) 2377.

\bibitem{BBmeson9}I. W. Lee, T. Lee, D. P. Min and B. Y. Park, Eur. Phys. J. {\bf C49}
(2007) 737.

\bibitem{BBmeson10}I. W. Lee and T. Lee, Phys. Rev. {\bf D76} (2007) 014017.

\bibitem{Simonov07} A. M. Badalian, Yu. A. Simonov and M. A. Trusov,
arXiv:0712.3943.

\bibitem{Matsuki05} T. Matsuki, K. Mawatari, T. Morii  and K. Sudoh, Phys. Lett. {\bf B606} (2005) 329.

\bibitem{Matsuki07} T. Matsuki, T. Morii and K. Sudoh, Prog. Theor. Phys. {\bf 117} (2007) 1077.


\bibitem{Wang0712} Z. G. Wang,  arXiv:0712.0118.

\bibitem{Wang0801} Z. G. Wang, arXiv:0801.1932.

\bibitem{Dashen} R. F. Dashen, Phys. Rev. {\bf 183} (1969) 1245.

\bibitem{Swanson06R} E. S. Swanson, Phys. Rept. {\bf 429} (2006) 243; and references  therein.

\bibitem{Colangelo04R} P. Colangelo, F. De Fazio and R. Ferrandes, Mod. Phys. Lett. {\bf
A19} (2004) 2083; and references  therein.

\bibitem{Wang23} Z. G. Wang and S. L. Wan, Phys. Rev. {\bf D73} (2006)
094020.

\bibitem{Wang24} Z. G. Wang, J. Phys. {\bf G34} (2007) 753.

\bibitem{Wang06} Z. G. Wang and S. L. Wan, Phys. Rev. {\bf D74} (2006) 014017.

\bibitem{Colangelo03} P. Colangelo and F. D. Fazio, Phys. Lett. {\bf B559} (2003)
49.

\bibitem{Wang04} Z. G. Wang, W. M. Yang and S. L. Wan,  Eur. Phys. J. {\bf C37}
(2004) 223.


\bibitem{LCSR1}
I. I. Balitsky, V. M. Braun and A. V. Kolesnichenko, Nucl. Phys.
{\bf B312} (1989) 509.

\bibitem{LCSR2}  V. L. Chernyak and I. R. Zhitnitsky, Nucl.
Phys. {\bf B345} (1990) 137.

\bibitem{LCSR3}V. L. Chernyak and A. R. Zhitnitsky, Phys. Rept. {\bf 112} (1984)
173.

\bibitem{LCSR4}V. M. Braun and I. E. Filyanov, Z. Phys.  {\bf C44} (1989)
157.

\bibitem{LCSR5}V. M. Braun and I. E. Filyanov, Z. Phys. {\bf C48} (1990) 239.

\bibitem{LCSRreview}
P. Colangelo and A. Khodjamirian, hep-ph/0010175.

\bibitem{SVZ791} M. A. Shifman, A. I. Vainshtein and V. I. Zakharov,
Nucl. Phys. {\bf B147} (1979) 385.

\bibitem{SVZ792} M. A. Shifman, A. I. Vainshtein and V. I. Zakharov,
Nucl. Phys. {\bf B147} (1979) 448.

\bibitem{Reinders85}  L. J. Reinders, H.
Rubinstein and S. Yazaki, Phys. Rept. {\bf 127} (1985) 1.



\bibitem{Narison89}  S. Narison, QCD Spectral Sum Rules, World Scientific Lecture
Notes in Physics {\bf 26} (1989) 1.

\bibitem{Belyaev94}
V. M. Belyaev, V. M. Braun, A. Khodjamirian and R. R\"uckl, Phys.
Rev. {\bf D51} (1995) 6177.

\bibitem{PSLC1} P. Ball, JHEP {\bf 9901} (1999) 010.

\bibitem{PSLC2} P. Ball and R. Zwicky, Phys. Lett. {\bf B633} (2006)
289.

\bibitem{PSLC3}  P. Ball and  R. Zwicky, JHEP {\bf 0602} (2006) 034.

\bibitem{PSLC4}  P. Ball, V. M. Braun and A. Lenz,
JHEP {\bf 0605} (2006) 004.



\bibitem{PDG} W.-M. Yao, et al, J. Phys. {\bf G33} (2006) 1.

\bibitem{Wang04BS} Z. G. Wang, W. M. Yang and S. L. Wan, Nucl. Phys. {\bf A744} (2004)
156.

\bibitem{Velasco07}J. M. Verde-Velasco, arXiv:0710.1790;  and references  therein.

\bibitem{HDress1} N. A. Tornqvist, Z. Phys. {\bf C68} (1995) 647.

\bibitem{HDress2} E. van Beveren and  G. Rupp, Phys. Rev. Lett. {\bf 91} (2003)
012003.

\bibitem{HDress3} Yu. A. Simonov and J. A. Tjon, Phys. Rev. {\bf D70} (2004) 114013.

\bibitem{Guo23} F. K. Guo, P. N. Shen,
H. C. Chiang and R. G. Ping, Phys. Lett. {\bf B641} (2006) 278.

\bibitem{Guo24} F. K. Guo, P. N. Shen and  H. C. Chiang, Phys. Lett. {\bf B647}
(2007) 133.


\bibitem{Zhu06} W. Wei, P. Z.  Huang and S. L. Zhu, Phys. Rev. {\bf D73} (2006) 034004.


\end{thebibliography}
\end{document}